\documentclass[pdflatex,sn-mathphys-num]{sn-jnl}

\usepackage{graphicx}%
\usepackage{multirow}%
\usepackage{amsmath,amssymb,amsfonts}%
\usepackage{amsthm}%
\usepackage{mathrsfs}%
\usepackage[title]{appendix}%
\usepackage{xcolor}%
\usepackage{textcomp}%
\usepackage{manyfoot}%
\usepackage{booktabs}%
\usepackage{algorithm}%
\usepackage{algorithmicx}%
\usepackage{algpseudocode}%
\usepackage{listings}%

\theoremstyle{thmstyleone}%
\theoremstyle{thmstyletwo}%

\theoremstyle{thmstylethree}%

\begin{document}

\title[Article Title]{Self-attention–enabled quantum path analysis of high-harmonic generation in solids}


\author*[1]{\fnm{Cong} \sur{Zhao}}\email{cong.zhao@kcl.ac.uk}
\equalcont{These authors contributed equally to this work.}
\author*[1]{\fnm{Xiaozhou} \sur{Zou}}\email{xiaozhou.1.zou@kcl.ac.uk}
\equalcont{These authors contributed equally to this work.}



\affil*[1]{Attosecond Quantum Physics Laboratory, King's College London, Strand, London, WC2R 2LS, United Kingdom}



\abstract{High-harmonic generation (HHG) in solids provides a powerful platform to probe ultrafast electron dynamics and interband–intraband coupling. However, disentangling the complex many-body contributions in the HHG spectrum remains challenging. Here we introduce a machine-learning approach based on a Transformer encoder to analyze and reconstruct HHG signals computed from a one-dimensional Kronig–Penney model. The self-attention mechanism inherently highlights correlations between temporal dipole dynamics and high-frequency spectral components, allowing us to identify signatures of nonadiabatic band coupling that are otherwise obscured in standard Fourier analysis. By combining attention maps with Gabor time–frequency analysis, we extract and amplify weak coupling channels that contribute to even-order harmonics and anomalous spectral features. Our results demonstrate that multi-head self-attention acts as a selective filter for strong-coupling events in the time domain, enabling a physics-informed interpretation of high-dimensional quantum dynamics. This work establishes Transformer-based attention as a versatile tool for solid-state strong-field physics, opening new possibilities for interpretable machine learning in attosecond spectroscopy and nonlinear photonics.}

\keywords{High-harmonic generation, Self-attention, Deep learning, Gabor analysis}

\maketitle

\section{Introduction}\label{sec1}

High-order harmonic generation (HHG) in solids has emerged as a powerful probe of ultrafast electron dynamics and light–matter interaction on the attosecond timescale \cite{cavaletto2025attoscience,heide2024ultrafast,goulielmakis2022high,ghimire2019high,cruz2024quantum}. 
In crystalline media, harmonic emission arises from two primary mechanisms: interband polarization~\cite{vampa2015semiclassical,vampa2017merge}, where an electron driven into the conduction band recombines with its associated hole in the valence band, and intraband current~\cite{vampa2015all,luu2016high,luu2015extreme}, where carriers undergo nonlinear acceleration within their respective bands. 
These two contributions interfere coherently, giving rise to the observed HHG spectrum. 
In particular, the concept of quantum paths—different trajectories taken by electron–hole pairs driven by a strong laser field—provides a fundamental framework for interpreting the microscopic origin of harmonic emission~\cite{zair2008quantum,salieres2001feynman}. 
However, this simplified decomposition does not capture all microscopic processes. 
In particular, when driven far from the adiabatic regime, electron wavefunctions can undergo rapid band-to-band transitions that lead to nonadiabatic coupling. 
This mechanism modifies the harmonic response beyond the standard picture~\cite{keldysh2024ionization,dykhne1962adiabatic,zener1932non,grifoni1998driven,sansone2004nonadiabatic}. A remarkable manifestation of this effect is the emergence of even-order harmonics in centrosymmetric systems, despite conventional symmetry arguments predicting their absence~\cite{luu2018measurement,uzan2020attosecond,uzan2022observation,uzan2024observation}.
Theoretical modeling provides a uniquely controlled platform for exploring these dynamics.
In this work, we employ a one-dimensional time-dependent Schrödinger equation (TDSE) simulation~\cite{l1992calculations,wu2015high,wu2016multilevel} using a Kronig–Penney potential~\cite{faisal1989exact,catoire2015above} to represent a periodic solid. 
This model encapsulates the essential physics of Bloch bands, tunneling, and field-driven electron–hole dynamics. 
Within this framework, we observe the clear occurrence of even-order harmonic components in the calculated spectra, arising directly from nonadiabatic interband coupling rather than structural asymmetry or extrinsic perturbations. 
This establishes the Kronig–Penney TDSE as an ideal minimal model for isolating and understanding nonadiabatic effects in solid-state HHG.
Recent advances in machine learning have unlocked new possibilities for extracting hidden structures within complex physical data~\cite{ha2021discovering,mototake2021interpretable,wetzel2020discovering,liu2021machine,rydzewski2023manifold}. 
Among these, the self-attention mechanism~\cite{vaswani2017attention,feng2024sliding,consens2025transformers}, originally developed for natural language processing, has proven particularly effective at capturing long-range correlations and enhancing relevant features in high-dimensional sequences. 
Applied to physical signals, self-attention offers a data-driven approach to highlight subtle couplings and disentangle overlapping modes without imposing restrictive prior assumptions.

To analyze these subtle contributions, we introduce a self-attention–based quantum path analysis. 
Central to transformer neural networks, the self-attention mechanism computes correlation weights between all pairs of times in the dipole response. 
Physically, these weights act as a nonlocal filter that amplifies strongly correlated two-time segments — specifically those where nonadiabatic transitions imprint phase asymmetry and symmetry breaking.
When applied to the TDSE dipole, the resulting attention map highlights temporal regions of enhanced coupling, while the reconstructed signals reveal the selective amplification of even-order harmonics.

By integrating the self-attention decomposition with Gabor time–frequency analysis~\cite{gabor1946theory,yao1993gabor}, we achieve a clear mapping from the raw TDSE dipole to the underlying nonadiabatic quantum paths.
This approach not only reproduces the full HHG spectrum but also isolates coupling-induced features that are otherwise masked in the dominant odd-order background. 

The results demonstrate how machine learning can be harnessed to uncover subtle strong-field processes, thereby bridging minimal quantum simulations with contemporary data-driven analysis.

This work establishes a new methodology for exploring solid-state HHG beyond the standard interband–intraband decomposition. 
By directly linking self-attention weights to nonadiabatic coupling signatures, we demonstrate that even in a minimal Kronig–Penney system, HHG encodes rich dynamical information that can be effectively extracted using appropriate analysis tools.
Our findings highlight the role of nonadiabaticity as a driver of symmetry breaking in HHG, and position self-attention as a promising technique for quantum path analysis in both theoretical and experimental contexts.

\section{Results}\label{sec2}

\subsection{Self-Attention Mechanism}
\begin{figure*}[ht!]
\centering\includegraphics[width=1\textwidth]{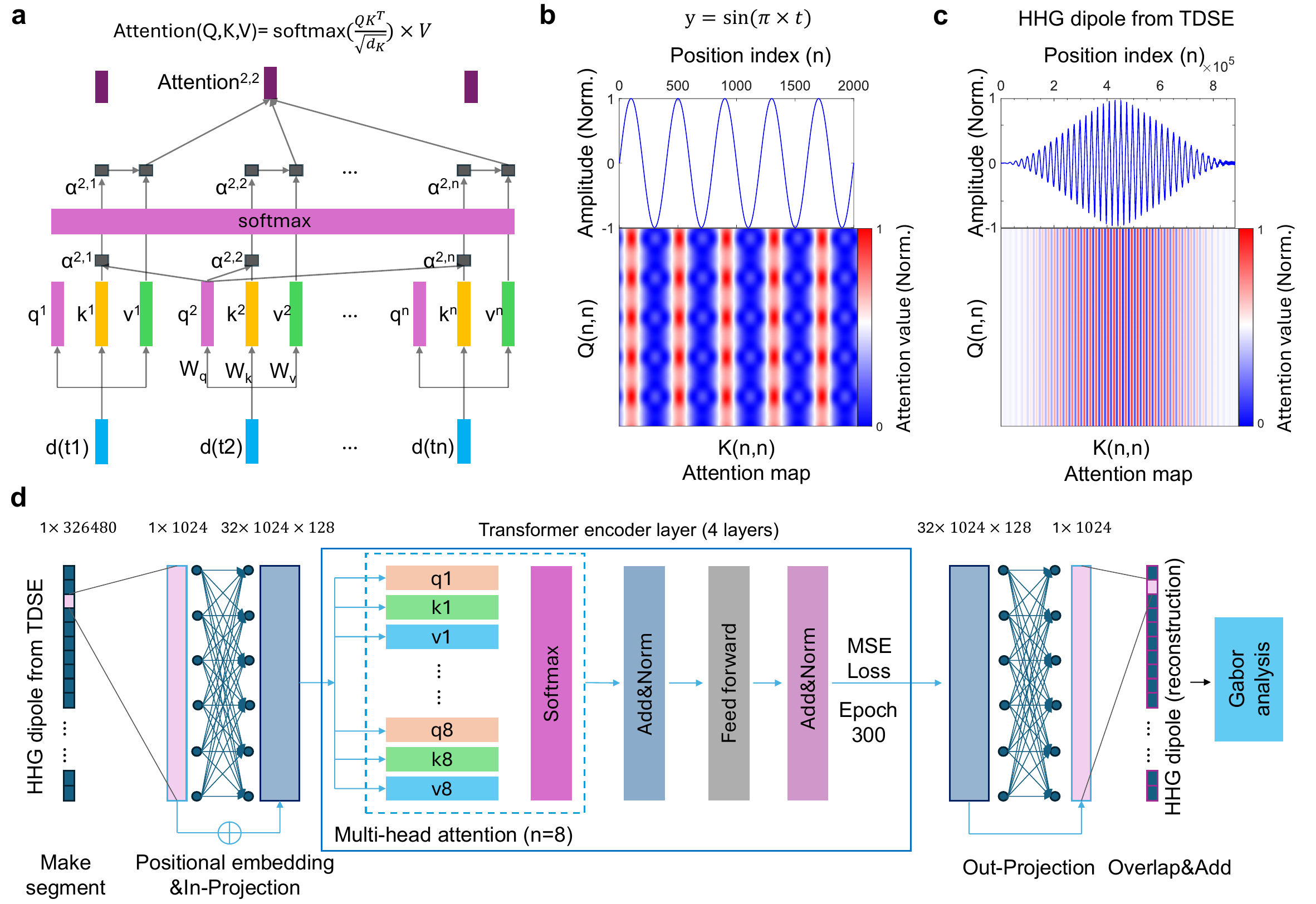}
\caption{\textbf{(a)}~Illustration of the self-attention mechanism in the HHG dipole analysis. Each $d(t_i)$ is extracted from the HHG dipole in the TDSE calculation. \textbf{(b)}~A simple sinusoid $y = \sin(\pi t)$ (top) produces an attention map (bottom) with regular red--blue bands, indicating that each time point attends to others with the same phase. \textbf{(c)}~HHG dipole from TDSE simulations (top) has fast oscillations with a slow envelope. Its attention map (bottom) shows extended vertical and horizontal correlations, capturing both short- and long-range temporal structures. \textbf{(d)}~The workflow of the Transformer-based analysis.}
\label{sketch}
\end{figure*}

In the Transformer framework, the central operation is the self-attention mechanism, which allows each element of the input sequence to adaptively aggregate information from all other elements. 
As shown in Fig.\ref{sketch}a, consider an input signal represented by a sequence of feature vectors:
\begin{equation}
   X = \begin{bmatrix} d(t_1)^\top \\ d(t_2)^\top \\ \vdots \\ d(t_n)^\top \end{bmatrix} \in \mathbb{R}^{n \times d_{\mathrm{model}}}, 
\end{equation}
where each vector $d(t_i)$ encodes the local representation at the $i$-th position of the HHG dipole. To construct the self-attention representation, the model first derives three sets of vectors—queries (Q), keys (K), and values (V)—through learned linear projections of the input:
\begin{equation}
   Q = X W_Q, \quad K = X W_K, \quad V = X W_V, 
\end{equation}
with $W_Q, W_K, W_V \in \mathbb{R}^{d_{\mathrm{model}} \times d_k}$. For each position $i$, the query vector $q_i = x_i W_Q$ specifies what information this element seeks, the key vector $k_i = x_i W_K$ specifies what information it provides, and the value vector $v_i = x_i W_V$ carries its content.  

The similarity between query $q_i$ and all keys $k_j$ is measured by a scaled dot product:
\begin{equation}
   \text{score}(i,j) = \frac{q_i k_j^\top}{\sqrt{d_k}}, 
\end{equation}
which is normalized using the softmax function to obtain attention weights
\begin{equation}
  \alpha_{ij} = \frac{\exp\!\left(\tfrac{q_i k_j^\top}{\sqrt{d_k}}\right)}{\sum_{j'=1}^n \exp\!\left(\tfrac{q_i k_{j'}^\top}{\sqrt{d_k}}\right)}.  
\end{equation}
The output attention value at position $i$ is then a weighted sum of value vectors:
\begin{equation}
   \text{attention}_i = \sum_{j=1}^n \alpha_{ij} v_j, 
\end{equation}
or, in compact matrix form,
\begin{equation}
   Attention = \text{softmax}\!\left(\frac{QK^\top}{\sqrt{d_k}}\right)V. 
\end{equation}
This operation allows the model to dynamically re-weight contributions from all positions, enabling it to capture long-range correlations and nonlocal patterns that are often inaccessible to traditional convolutional or recurrent architectures.  

In the present context, the input sequence $Attention$ corresponds to the time-dependent dipole response obtained from TDSE simulations of high-harmonic generation in solids (see Methods for more details). Each $d(t_i)$ represents the local dipole signal at time step $t_i$, enriched by positional encoding to retain temporal ordering. 
Through self-attention, the model can explicitly learn how different temporal segments of the dipole interact—such as nonadiabatic coupling between electronic bands or the delayed buildup of even-order harmonics—thereby extracting physical features that are directly relevant to the interpretation of HHG spectra.

Fig.\ref{sketch}b and Fig.\ref{sketch}c demonstrate how the self-attention mechanism responds to signals of different complexity. In fig.\ref{sketch}b, the input is a simple sinusoidal function $y=\mathrm{sin}(\pi t)$ (top), where the periodic oscillation leads to an attention map (bottom) with sharp, regularly spaced red–blue bands along both the Q and K axes.
This pattern indicates that each time point strongly attends to other points with the same phase in the oscillation, faithfully reproducing the periodic structure of the input. 
In Fig.\ref{sketch}c, the input is the HHG dipole moment obtained from TDSE simulations (top), which features rapidly oscillating components modulated by a slowly varying envelope. 
The resulting attention map (bottom) is no longer strictly periodic but instead exhibits extended vertical and horizontal correlation structures, reflecting the coexistence of short-range carrier oscillations and long-range envelope dynamics. 
This highlights the ability of attention to disentangle and represent both local and global temporal correlations in HHG signals.

The workflow of our Transformer-based analysis is summarized in Fig.\ref{sketch}d. 
The HHG dipole signal obtained from TDSE simulations is first segmented into shorter windows, which are then embedded into a higher-dimensional representation through positional encoding and an input projection layer. 
These embedded sequences are processed by a four-layer Transformer encoder, where multi-head self-attention (with eight heads) computes correlations between different time positions of the signal. 
Each encoder layer includes the standard sequence of operations: multi-head attention, residual connection with layer normalization, and a feed-forward network. 
The model is trained for 300 epochs by minimizing the mean squared error (MSE) between the reconstructed and original dipole segments. 
After passing through the encoder, the features are projected back to the time domain using an output projection followed by an overlap-and-add procedure to reconstruct the full HHG dipole signal. 
The reconstructed signal is then analyzed using a Gabor transform to extract its spectral–temporal features, allowing us to directly compare the learned attention patterns with conventional time–frequency analysis.

\subsection{Reconstructed dipole after Transformer-based analysis}

\begin{figure*}[ht!]
\centering\includegraphics[width=1\textwidth]{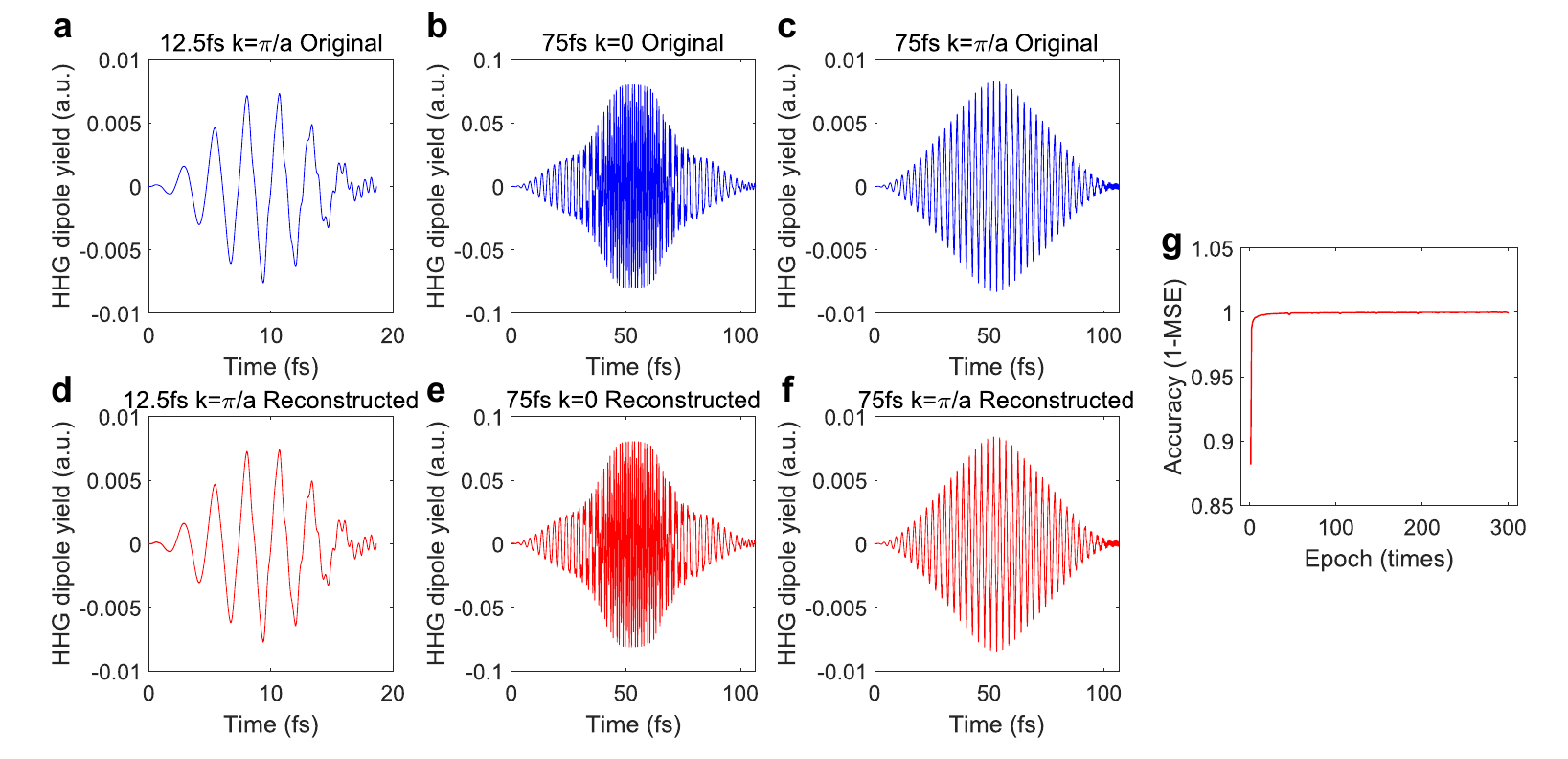}
\caption{Transformer reconstruction against TDSE dipole signals. (\textbf{a}--\textbf{c}) Representative HHG dipoles from TDSE simulations at different propagation times and crystal momenta: \textbf{(a)} 12.5~fs segment at $k=\pi/a$, \textbf{(b)} 75~fs signal at $k=0$, and \textbf{(c)} 75~fs signal at $k=\pi/a$. (\textbf{(d)}--\textbf{(f)}) Corresponding Transformer reconstructions, capturing both the oscillatory structure and envelope modulation, demonstrate retention of short-time dynamics and long-time coherence. Minor deviations appear in fine details, especially for the short segment, but the overall waveform is preserved. \textbf{(g)} Training curve showing the reconstruction accuracy ($1-\text{MSE}$) converging rapidly and stabilizing near unity after 300 epochs, indicating high-fidelity performance across different conditions.}
\label{fig1}
\end{figure*}

To validate the performance of the Transformer framework, we compare the reconstructed dipoles with the original TDSE signals under different conditions, as shown in Fig.~\ref{fig1}. 
The original dipoles for short (12.5~fs, FWHM in intensity) and long (75~fs) propagation times at both $k=0$ and $k=\pi/a$ are presented in panels (a--c), while the corresponding reconstructions are shown in panels (d--f). 
The Transformer successfully reproduces the key temporal features of the dipole dynamics, including the rapid carrier oscillations and the slowly varying envelope, demonstrating its ability to capture both short-time and long-time correlations. 
Although small deviations appear in the fine structures of the shorter segment, the overall waveform is well preserved. The training convergence, shown in Fig.~\ref{fig1}g, further confirms the model’s robustness: the accuracy (defined as $1-\mathrm{MSE}$) rises quickly within the first few epochs and saturates close to unity after 300 epochs, indicating that the model achieves high-fidelity reconstruction across different temporal regimes and crystal momenta.

\subsection{Non-adiabatic coupling extracted by self-Attention Mechanism}
\begin{figure*}[ht!]
\centering\includegraphics[width=1\textwidth]{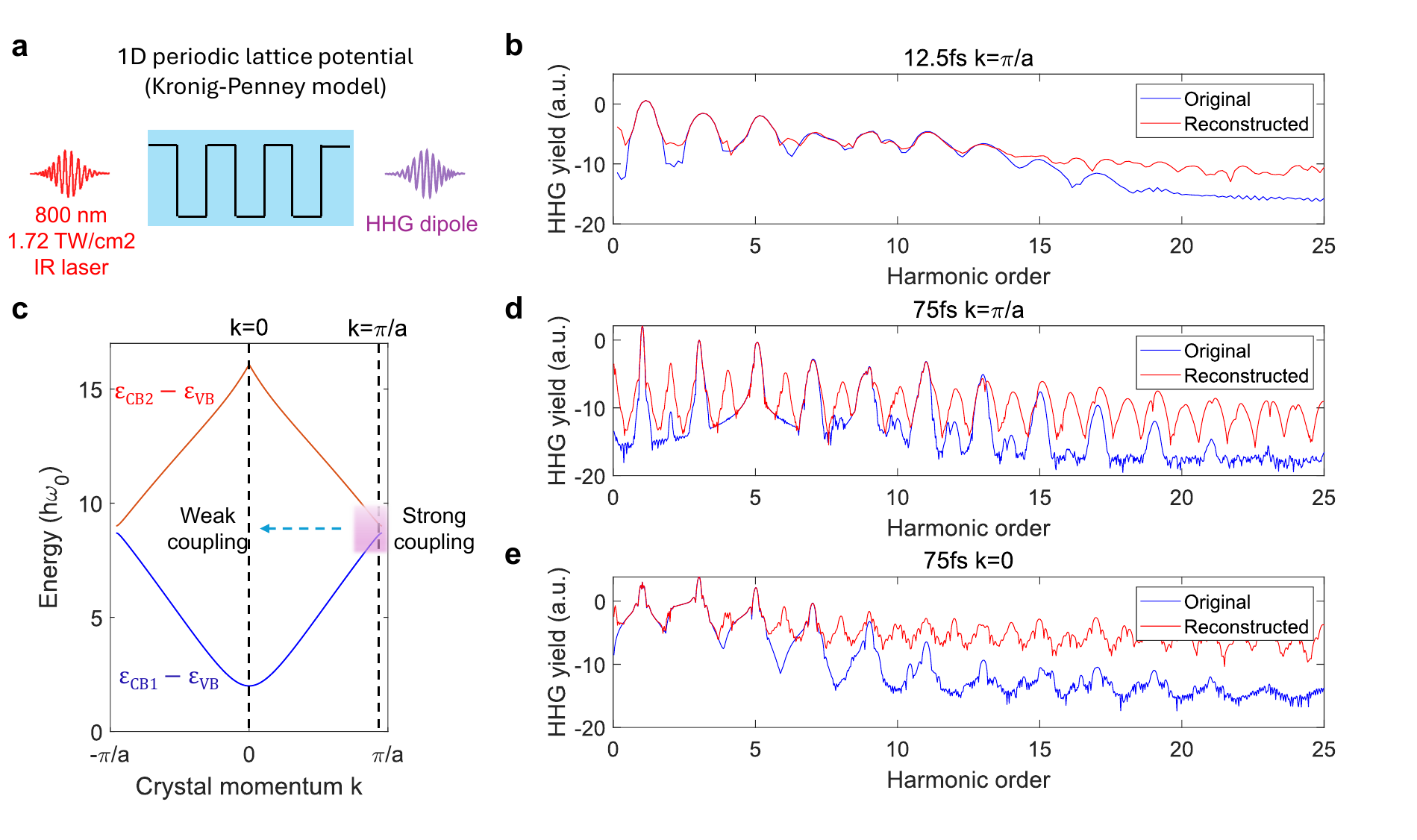}
\caption{Comparison between original TDSE and Transformer-reconstructed HHG spectra. \textbf{(a)}~Schematic of the simulation setup based on a one-dimensional Kronig–Penney lattice potential driven by a strong infrared laser. \textbf{(b,d,e)}~Representative harmonic spectra at different propagation times and crystal momenta, showing that the Transformer faithfully reproduces the main features of the TDSE results, including the plateau and cutoff. The even harmonics produced by the long pulse (75 fs) indicate the nonadiabatic coupling, and the reconstructed HHG spectrum can capture and amplify this dynamic. \textbf{(c)}~Bandgap structure indicating valence band (VB), first conduction band (CB1), and second conduction band (CB2); strong non-adiabatic coupling emerges near the Brillouin-zone edge ($k=\pi/a$), where the energy gap is small. The excellent agreement, especially at $k=\pi/a$, demonstrates that the self-attention mechanism captures spectral signatures of interband tunneling and band coupling, while only minor deviations appear at low-order harmonics.}
\label{fig2}
\end{figure*}

An important outcome of applying the self-attention mechanism to high-harmonic generation (HHG) signals is its capacity to uncover signatures of nonadiabatic coupling between electronic states. In conventional semiconductor HHG theory, interband and intraband contributions are usually distinguished through analytical models. However, when electronic wave packets are driven to the edge of the Brillouin zone, adjacent conduction bands may be separated by less than a single photon energy. This proximity gives rise to complex dynamics~\cite{lignier2007dynamical}, including Landau–Zener transitions and nonadiabatic band coupling. Directly resolving the interplay of these processes in the time domain, however, remains a major challenge. 

The attention maps obtained from the Transformer provide a data-driven route to visualize these couplings: instead of showing only local periodic correlations, the maps display extended, off-diagonal structures that encode long-range temporal dependencies in the dipole response. These features arise from the breakdown of adiabatic electron dynamics, where the carrier rapidly transitions between different bands under strong-field driving. By learning these correlations directly from the TDSE dipole, the self-attention mechanism effectively extracts non-adiabatic coupling as an emergent pattern in the data, offering a powerful alternative to traditional perturbative analyses.

To assess the reliability of our Transformer framework at the spectral level, we reconstruct the HHG spectra against the original TDSE results, as shown in Fig.~\ref{fig2}. The simulation setup is illustrated in Fig.~\ref{fig2}a, where a one-dimensional Kronig--Penney lattice potential is driven by a strong infrared laser pulse to generate the HHG dipole response.  

The corresponding bandgap structure is presented in Fig.~\ref{fig2}c. The first two energy gaps, $\epsilon_{CB1}-\epsilon_{VB}$ and $\epsilon_{CB2}-\epsilon_{VB}$, define the main channels for strong-field dynamics. Near the Brillouin zone center ($k=0$), the two energy gaps are relatively large, leading to weak interband mixing and predominantly adiabatic dynamics. By contrast, close to the zone boundary ($k=\pi/a$), the energy gaps become small, and strong non-adiabatic coupling between bands emerges, which plays a critical role in shaping the HHG spectrum.  

Fig.~\ref{fig2}b,d,e compare the original (blue) and Transformer-reconstructed (red) spectra under different conditions. At $k=\pi/a$, both short (12.5~fs, Fig.~\ref{fig2}b) and long (75~fs, Fig.~\ref{fig2}d) propagation times exhibit good agreement in the plateau region and harmonic cutoffs. Interestingly, the Transformer reconstruction at long propagation time (Fig.~\ref{fig2}d) reveals pronounced even-order harmonics that are much weaker in the original TDSE spectrum. A similar effect is observed at $k=0$ for the long propagation case (Fig.~\ref{fig2}e). The appearance of strong even-order harmonics indicates that the Transformer has effectively amplified the role of non-adiabatic coupling, which breaks the inversion symmetry of the electron dynamics and allows otherwise suppressed even-order processes to emerge.  

This spectral behavior is consistent with the attention maps discussed earlier in Fig.~\ref{sketch}c, where off-diagonal structures encoded long-range temporal correlations beyond simple periodicity. Those correlations arise from rapid interband transitions, i.e., non-adiabatic coupling, and are here manifested in the form of enhanced even-order harmonics. Together, the attention-domain analysis and the spectral reconstructions demonstrate that the Transformer framework not only preserves the main spectral features of the HHG signal but also strengthens the fingerprints of non-adiabatic band mixing, providing a complementary perspective to conventional time--frequency methods.

\subsection{Gabor analysis revealing non-adiabatic coupling}
\begin{figure*}[ht!]
\centering\includegraphics[width=1\textwidth]{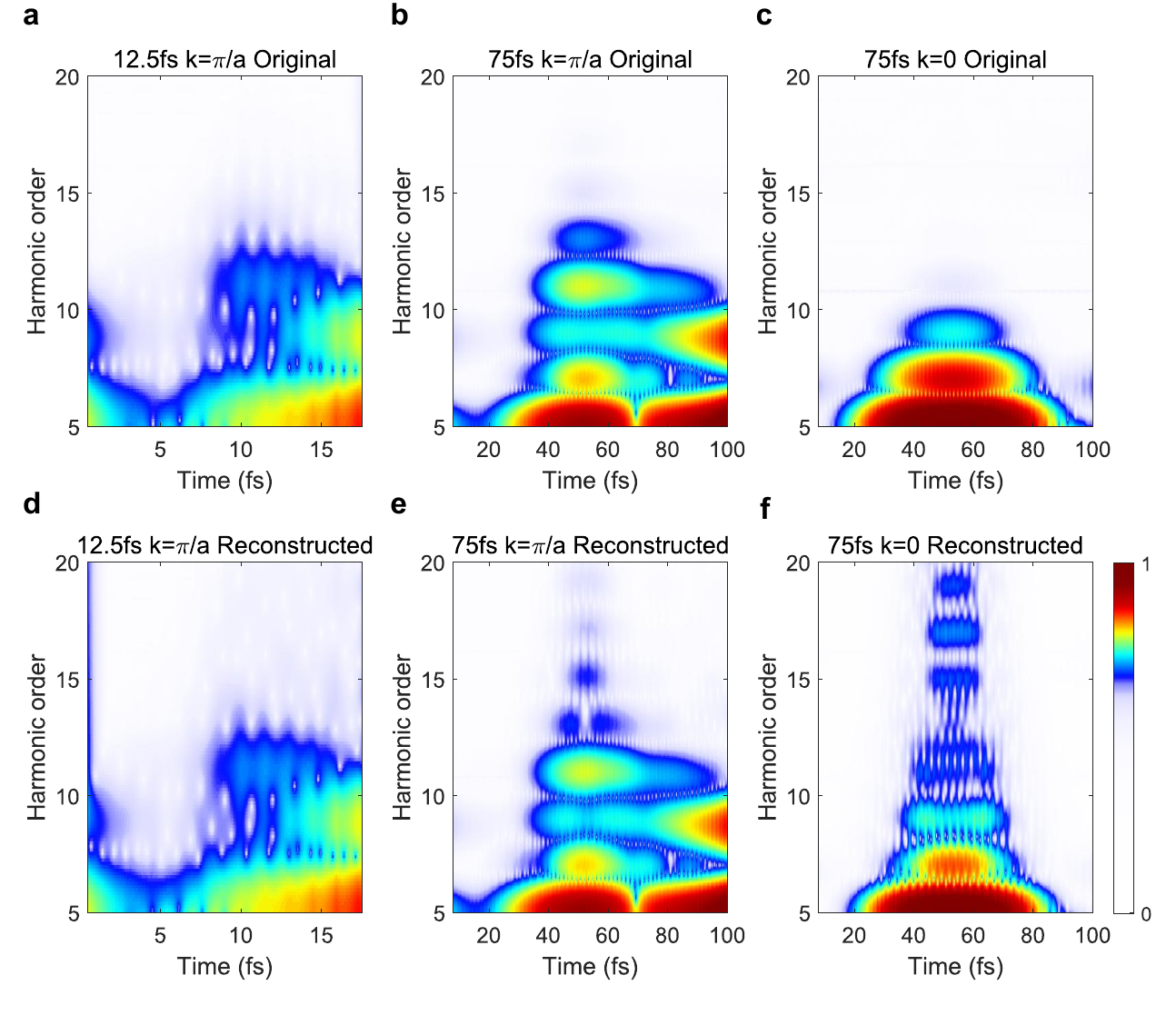}
\caption{Comparison of original and Transformer-reconstructed HHG Gabor spectra.\textbf{a-c} present Gabor time–frequency spectra of high-harmonic emission obtained from TDSE simulations: \textbf{a} a 12.5 fs segment at $k=\pi/a$, \textbf{b} a 75 fs signal at $k=\pi/a$, and \textbf{c} a 75 fs signal at $k=0$. \textbf{d-f} show the corresponding Transformer-reconstructed spectra under the same conditions. \textbf{e} and \textbf{f} reveal additional features beyond the tenth harmonic, where elliptical regions indicate the presence of coupled electronic states and vertical blue lines correspond to abrupt non-adiabatic transitions between them. At
$k=\pi/a$ (\textbf{e}), the ellipses are broader and more extended, reflecting stronger state mixing and enhanced non-adiabatic coupling at the Brillouin-zone edge, while at $k=0$ (\textbf{f}) the transitions
are sharper and more localized in time. These observations demonstrate that the Transformer not only reconstructs the HHG dipoles but also captures hidden signatures of non-adiabatic dynamics in solids. The color scale denotes normalized intensity.}
\label{fig3}
\end{figure*}

To further investigate the role of non-adiabatic coupling in solid-state HHG, we perform a Gabor time–frequency analysis of the dipole signals, as summarized in Fig.~\ref{fig3}. Fig.~\ref{fig3}a-Fig.~\ref{fig3}c present the spectra directly obtained from TDSE simulations for different crystal momenta and propagation times. At $k=\pi/a$, a short 12.5~fs segment (Fig.~\ref{fig3}a) primarily displays odd-order harmonics with a smooth temporal modulation. When extending the propagation window to 75~fs (Fig.~\ref{fig3}b), the harmonic structure remains dominated by odd orders, consistent with the expectation from inversion-symmetric crystals in the adiabatic limit. Similarly, the spectrum at $k=0$ (Fig.~\ref{fig3}c) shows only weak signatures of even-order contributions, confirming that the direct TDSE response is largely governed by interband and intraband dynamics under nearly adiabatic conditions.  

Fig.~\ref{fig3}d-Fig.~\ref{fig3}f display the corresponding results reconstructed using the Transformer model. For the short-time segment at $k=\pi/a$ (Fig.~\ref{fig3}d), the reconstructed spectrum closely reproduces the TDSE benchmark, establishing the accuracy of the model in preserving harmonic structures. However, for the longer propagation window at $k=\pi/a$ (Fig.~\ref{fig3}e), the Transformer output exhibits pronounced even-order harmonics that are nearly absent in the original TDSE spectrum. A similar effect is observed at $k=0$ (Fig.~\ref{fig3}f), where strong even orders emerge in addition to the odd harmonics. This systematic enhancement of even-order signals highlights the ability of the self-attention mechanism to amplify subtle dynamical correlations that are otherwise buried in the raw TDSE response.  

A closer inspection of the reconstructed spectra in Fig.~\ref{fig3}e and Fig.~\ref{fig3}f reveals additional structures that are absent in the TDSE benchmarks. Beyond the tenth harmonic, we observe the appearance of elliptical features together with nearly vertical blue lines. The elliptical areas can be attributed to the formation of coupled states induced by strong-field driving, which is a hallmark of enhanced non-adiabatic coupling. These features indicate that the Transformer reconstruction can amplify correlations that correspond to electronic states undergoing dynamical mixing rather than remaining confined to adiabatic trajectories.  

The vertical blue lines, on the other hand, signify abrupt non-adiabatic transitions between different coupled states. Their presence suggests that the Transformer uncovers temporal channels of population transfer that are otherwise weak in the raw TDSE response. Interestingly, while both $k=\pi/a$ (Fig.~\ref{fig3}e) and $k=0$ (Fig.~\ref{fig3}f) exhibit these non-adiabatic signatures, the details differ: in Fig.~\ref{fig3}e, the elliptical regions are more pronounced and extend over a broader frequency range, reflecting stronger state mixing at the Brillouin-zone edge. In contrast, Fig.~\ref{fig3}f shows sharper and more isolated vertical transitions, indicating that at the zone center, non-adiabatic effects are more localized in time.  

These observations reinforce the conclusion that the Transformer does more than reproduce the TDSE spectra: it systematically enhances hidden non-adiabatic channels, thereby exposing both the coupling of electronic states (elliptical features) and their transition dynamics (vertical lines). This dual signature demonstrates that self-attention can serve as a sensitive probe of dynamical symmetry breaking and electronic correlation in solid-state HHG.

The appearance of strong even-order harmonics in the reconstructed spectra provides direct evidence of strengthened non-adiabatic coupling. In particular, the Transformer appears to capture correlations between interband polarization and intraband motion that break inversion symmetry dynamically, giving rise to even-order responses. Thus, the self-attention framework does not merely interpolate the original data, but instead extracts hidden features of the underlying electronic dynamics. This result demonstrates the potential of Transformer-based analysis as a diagnostic tool for probing non-adiabaticity in strong-field solid-state HHG.

\section{Conclusion}\label{sec13}
In this work, we have demonstrated that Transformer-based models provide a powerful framework for analyzing high-harmonic generation in solids. 
By reconstructing time-dependent dipole responses from TDSE simulations, the self-attention mechanism enables direct identification of non-adiabatic coupling effects in the harmonic spectra. 
In particular, we showed that the Transformer not only accurately reproduces the characteristic the plateau and cutoff structures of the spectral, but also enhances subtle features such as even-order harmonics and vertical electronic transitions. 
These spectral signatures show the presence of strong non-adiabatic interactions between electronic bands.

Our results highlight the potential of attention-based architectures as data-driven tools for probing ultrafast electron dynamics, offering insights that complement to traditional semiclassical or band-structure analyses.

Beyond the specific Kronig–Penney model considered here, this methodology can be generalized to more realistic materials and experimental data. 
It provides a versatile platform for disentangling interband and intraband contributions.
In addition, it enables mapping coupled electronic states, and exploring light–matter interactions in complex solids. 
We expect that the combination of advanced machine learning models with attosecond strong-field physics will open new opportunities for quantitatively characterizing non-adiabatic processes and guiding the design of novel photonic and electronic materials. 

\section{Method}\label{secA1}

\subsection{TDSE calculation}

We numerically solve TDSE with a 1D model potential $V(x)$, which follows the Kronig-Penney model. In this model, the potential well is centred at $x = 0$ and has a width of $a/2$ with $a = 8.2~\text{a.u.}$ being the periodicity of the system in the real space.

The wave function describing the system is given by the Bloch wave function $\Psi(x,t) = e^{ik_0x}u(k_0,x,t)$ with the reduced wave function u($k_0$,x,t), which is periodic in space with $2\pi/a$ periodicity, and the initial momentum $k_0$. The reduced part of the Bloch wave function is obtained by solving the TDSE in the dipole approximation in the velocity gauge. The TDSE is explicitly given by
\begin{equation}
    i\frac{\partial u}{\partial t} = \tilde H(k_0,t)u = \left\lbrace\frac{[P+k_0+A(t)]^2}{2}+V(x)\right\rbrace u \, ,
\end{equation}
where $P$ is the electron momentum operator. $A(t)$ is the vector potential of the electromagnetic field, which has a central frequency $\omega_0$ and an amplitude $A_0 = E_0/\omega_0$. Here $\omega_0=0.057~\text{a.u.}$ (corresponding to wavelength $800~\text{nm}$) and $E_0=0.007~\text{a.u.}$.
Using the velocity gauge ensures that we keep the periodicity of the Hamiltonian over time. 

The wave function is decomposed on a plane wave basis and is solved directly in real space. The propagation is performed using the Lanczos algorithm (with a basis of size $10$) with the initial condition $u(k_0,x,t = 0) = \varphi(k_0,x)$. $\varphi(k_0,x)$ is the eigenstate of the stationary Schr\"odinger equation with eigenvalue $\epsilon_\mathrm{n}(k_0)$, describing the band $n$. In this work, $\varphi(k_0,x)$ is the eigenstate for $n\geq 1$. 
Here we start with $n = 2$ to get the bandgap at the $\mathrm{\Gamma}$ point. The numerical convergence is checked using the procedure of varying the grid parameters. Typically, 21 plane waves are used to fully converge the TDSE solution. The calculations are performed in the velocity gauge to ensure momentum conservation.

Once the wave function over time has been obtained, the HHG spectrum is calculated from the derivative of the current using the Ehrenfest theorem:
\begin{equation}
    \frac{\partial J(k_0,t)}{\partial t} =  \int dx \, u^* (k_0,x,t) 
    \left[ \frac{\partial V}{\partial x} + E(t) \right] u(k_0,x,t) \, .
\end{equation}
The total current is then found by integrating over $k_0$ in the BZ, that is 
\begin{equation}
\frac{\partial J(t)}{\partial t} =\int_{BZ} dk_0 \, \frac{\partial J(k_0,t)}{\partial t} \, .
\end{equation}
The power spectrum of the emitted harmonics is then evaluated by taking the Fourier transform of the current as
\begin{equation}
    I_\mathrm{HHG}(\omega)=\left| \omega\int_{-\infty}^{\infty}dt \, e^{i\omega t}J(t) \right|^2 \, .
\end{equation}

\subsection{Gabor Analysis}

To resolve the temporal evolution of high-harmonic generation (HHG) spectra, we employed Gabor analysis, which provides a joint time–frequency representation of the dipole signal. The Gabor transform is defined as a windowed Fourier transform, in which the signal $d(t)$ is multiplied by a Gaussian envelope centered at time $t_0$ and then Fourier transformed, yielding
\begin{equation}
G(\omega, t_0) = \int_{-\infty}^{\infty} d(t)\, 
\exp\!\left[-\frac{(t-t_0)^2}{2\sigma^2}\right] 
e^{-i\omega t}\, dt ,
\end{equation}
where $\sigma$ controls the width of the temporal window. This representation balances time and frequency resolution, enabling us to visualize the emission time of different harmonic orders.

In our calculations, the dipole acceleration obtained from time-dependent Schrödinger equation (TDSE) simulations was used as the input signal $d(t)$. The window width $\sigma$ was chosen to optimize the trade-off between temporal localization and spectral resolution, ensuring that both sub-cycle dynamics and plateau harmonics could be resolved. The resulting $|G(\omega,t_0)|^2$ maps were used to analyze the temporal buildup of harmonic radiation and to identify signatures of non-adiabatic transitions. In particular, elliptical regions and vertical streaks in the Gabor maps correspond to coupled states and sudden interband transitions, respectively, providing a direct probe of non-adiabatic coupling during HHG.

\section{Acknowledgments}
This work is supported by Royal Society funding via A.Z.'s research project ‘AMOS’~RGS\textbackslash{}R1\textbackslash{}211053. Computer time for this study was provided by the computing facilities of the MCIA~(Mésocentre de Calcul Intensif Aquitain) and King's Computational Research, Engineering and Technology Environment~(CREATE) from King's College London.

\bibliography{sn-bibliography}

\end{document}